\documentclass[preprint]{revtex4-1}

\draft % marks overfull lines with a black rule on the right

\usepackage{graphicx}% Include figure files
\usepackage{dcolumn}% Align table columns on decimal point
\usepackage{bm}% bold math

\begin{document}

%\twocolumn[ %% activate for two-column option

\title{Cascaded four-wave mixing in tapered plasmonic nanoantenna}

%% For REVTeX it is possible to automate superscript and e-mail callouts with the superscriptaddress option; see REVTeX4 documentation.

\author{I. S. Maksymov,$^{*}$ A. E. Miroshnichenko, and Yu. S. Kivshar}

\address{
Nonlinear Physics Centre, Australian National University, Research School of Physics and Engineering,\\ Canberra ACT 0200, Australia \\
$^*$Corresponding author: mis124@physics.anu.edu.au
}

\begin{abstract}
\noindent We study theoretically the cascaded four-wave mixing (FWM) in broadband tapered plasmonic nanoantennas and demonstrate a $300$-fold increase in nonlinear frequency conversion detected in the main lobe of the nanoantenna far-field pattern. This is achieved by tuning the elements of the nanoantenna to resonate frequencies involved into the FWM interaction. Our findings have a potentially broad application in ultrafast nonlinear spectroscopy, sensing, on-chip optical frequency conversion, nonlinear optical metamaterials and photon sources.
\end{abstract}

\maketitle %\maketitle must follow title, authors, abstract and \pacs

\noindent Ultrasmall plasmonic nanoantennas have recently attracted considerable attention because of their unique capability to confine light at the subwavelength scale (see, e.g., \cite{first_novotny_rev, review}). Nonlinear optical effects, such as second, and higher, harmonic generation and four-wave mixing (FWM), can be enhanced in a controllable way using light focusing properties of plasmonic nanoantennas \cite{lip05, dan07, pal08, gra10, ko11, sch11, har12, sla12, abb12, cia12, hen12}. This enhancement is of utmost importance for a variety of applications ranging from sensing \cite{ko11}, ultrafast nonlinear spectroscopy \cite{sch11}, on-chip optical frequency conversion \cite{hen12} to the design of metamaterials with enhanced optical response \cite{nl_metamater} and control of photon sources \cite{cia12, hen12, mak_oe_12}. However, most nonlinear nanoantennas reported in the literature so far exhibit a narrowband response because of their dipolar nature \cite{first_novotny_rev}, and consequently they are not suitable for efficient control of both fundamental and generated waves by cascaded FWM process. Cascaded FWM is observed in relatively long optical fibers and is extremely useful to build compact multiline spectral sources for dense wavelength division multiplexing (DWDM) as well as for the generation of short optical pulses or frequency combs for metrology \cite{agrawal}.

In this Letter, we demonstrate theoretically a substantial enhancement of cascaded FWM achieved with a broadband tapered plasmonic nanoantenna \cite{apl11,oc11,pav12,cia12} embedded in a nonlinear medium. By properly choosing the taper geometry of a realistic architecture \cite{review} we satisfy phase-matching conditions essential for strong FWM \cite{agrawal} and also additionally boost the frequency conversion efficiency using localized plasmon resonances. 

\begin{figure}[htb]
\centerline{
\includegraphics[width=8.5cm]{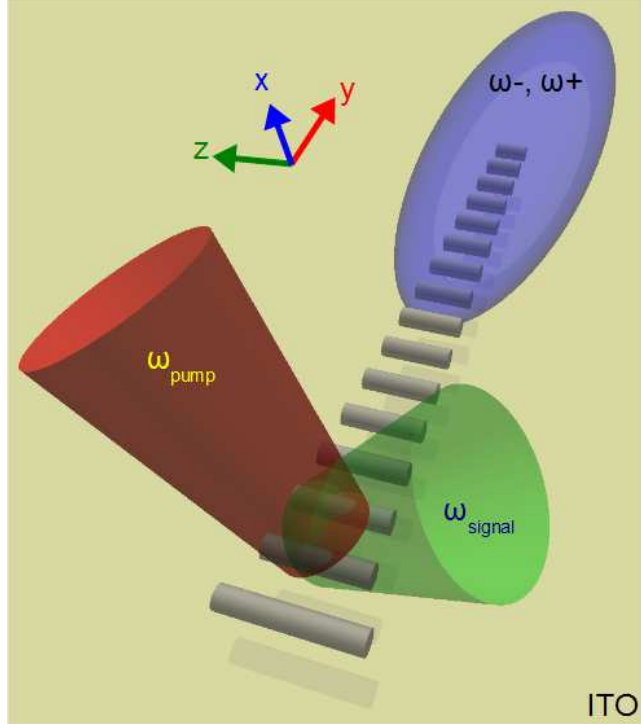}}
\caption{Illustration of the cascaded FWM conversion with a tapered nanoantenna. Two fields, pump and signal, with different telecom frequencies excite the nanoantenna surrounded by a nonlinear medium (ITO). Nanoantenna supports up to $10$ frequencies on both sides of the pump. Waves at these frequencies are emitted by the nanoantenna into a narrow cone in the \textit{y}-direction.}
\end{figure}

\begin{figure*}[htb]
\centerline{
\includegraphics[width=18cm]{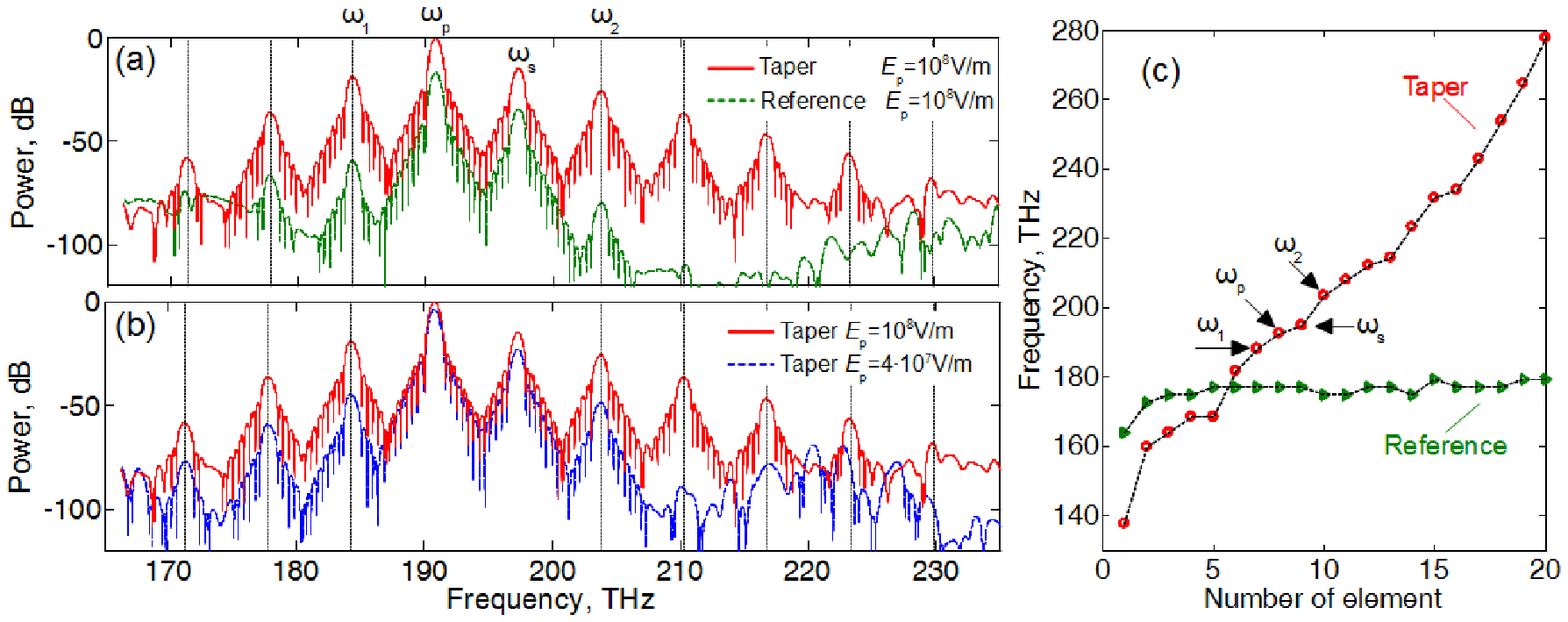}}
\caption{(a, b) Far-field power spectra of the $21$-element tapered nanoantenna (red solid line: $E_{\rm{p}} = 10^8$ V/m; blue dashed line: $4 \cdot 10^7$ V/m) and reference structure with each element tuned to $\omega_{\rm{p}}$ when alone (green dashed line, $10^8$ V/m). Straight dotted lines indicate frequencies generated due to FWM. All spectra are normalized to the peak value of the red curve. (c) Near-field resonance frequencies of the individual elements of the \emph{linear} tapered nanoantenna (red circles) and reference structure (green triangles) excited with a plane wave incident from the far-field region in the $-y$ direction. Here [and also in Fig. $3$(a)] the black dashed lines are the guides to the eye only.}
\end{figure*}

FWM is a nonlinear process transferring energy from a strong pump wave to waves upshifted and downshifted in frequency from the pump frequency \cite{agrawal}. When a strong pump wave at the frequency $\omega_{\rm{p}}$ and a weak signal wave at the frequency $\omega_{\rm{s}}$ ($\omega_{\rm{s}}>\omega_{\rm{p}}$) propagate in a nonlinear medium, photons from these waves annihilate and new photons are created at the frequencies $\omega_{\rm{1}} = 2\omega_{\rm{p}} - \omega_{\rm{s}}$ and $\omega_{\rm{2}} = 2\omega_{\rm{s}} - \omega_{\rm{p}}$. Furthermore, the created waves participate further in the FWM process leading to the cascaded generation of waves at frequencies around $\omega_{\rm{s}}$ and $\omega_{\rm{p}}$ 
\begin{eqnarray}
\omega_{\rm{-}}  = 2\omega_{\rm{low}} - \omega_{\rm{high}},~~~\omega_{\rm{+}} = 2\omega_{\rm{high}} - \omega_{\rm{low}}
\label{eq:one},
\end{eqnarray}
where $\omega_{\rm{low}}$ and $\omega_{\rm{high}}$ denote generated frequencies belonging to the low- and high-frequency sidebands.  

Significant FWM occurs only if the phase mismatch nearly vanishes, which can be satisfied when the frequencies involved into FWM are close to each other \cite{agrawal}. We meet these conditions and, moreover, further boost the strength of cascaded FWM using a nonlinear tapered nanoantenna (Fig. $1$). We embed the nanoantenna in Indium-Tin-Oxide (ITO) background and also gradually decrease the length of its silver elements in accord with Eq.($1$) and optical properties of ITO. ITO has the dielectric constant $\epsilon = 2.89$ and third-order susceptibility $\chi^{(3)} = 2.16 \cdot 10^{-18}$ m$^2$/V$^2$ \cite{hum06}, giving rise to an instantaneous nonlinear response. We assume that $\chi^{(3)}$ is constant within the operating band of the nanoantenna and also neglect the nonlinear response of silver because the exciting field penetrates only weakly into the metal and is mostly concentrated between the metal parts, and, therefore, predominantly interacts with ITO. The tuning of the operating band of the nanoantenna to the near-IR spectral region is achieved by rescaling the length of its elements as $L_{\rm{ITO}} = L_{\rm{air}}/\sqrt{\epsilon_{\rm{ITO}}}$, where $L_{\rm{air}}$ is the length of the elements of the same nanoantenna but suspended in air. The choice of the rest of the parameters is dictated by the resolution attainable with the current fabrication technologies \cite{review}: The elements of the nanoantenna and reference structure have the square cross-section of $50$ nm per $50$ nm and are separated by the spacing of $30$ nm. Experimental data for frequency-dependent dielectric permittivity of silver are taken from \cite{palik}. 

Using a three-dimensional finite-difference time-domain (FDTD) method \cite{fdtd1,fdtd2}, we investigate \emph{nonlinear} characteristics of the tapered nanoantenna by exciting it with the pump wave at frequency $190.8$ THz and the signal wave at $197.3$ THz. For the sake of illustration only, we use two amplitudes of the pump wave: $E_{\rm{p}} = 4 \cdot 10^7$ V/m and $10^8$ V/m  causing, correspondingly, nonlinear refractive index change of $\sim 10^{-3}$ and $\sim 5 \cdot 10^{-2}$ in bulk material. The amplitude of the signal wave equals to $0.1E_{\rm{p}}$. We simulate the far-field response of the nanoantenna by integrating the power flux through a plane perpendicular to the $y$-axis and located far beyond the nanoantenna front end. 

For the $21$-element nanoantenna and $E_{\rm{p}} = 10^8$ V/m [Fig. $2$(a)] we observe the signal generation due to FWM in accord with Eq.($1$). The tapered nanoantenna enhances the pump and signal waves as well as four generated waves above $-50$ dB in the low- and high-frequency bands. The enhancement is quantified by the frequency conversion efficiency calculated as $\eta = |P_{\rm{c}}|/|P_{\rm{s}}|$, where $P_{\rm{c}}$ is the power of the converted wave at $\omega_{\rm{-,+}}$ and $P_{\rm{s}}$ is the power of the signal wave in the detected far-field. For example, for the converted waves at $184.3$ THz and $203.8$ THz we obtain $\eta = -4.2$ dB and $-11.2$ dB, correspondingly. A qualitatively similar picture is observed for the $21$-element nanoantenna but excited with a weaker pump wave [Fig. $2$(b)]: for $E_{\rm{p}} = 4 \cdot 10^7$ V/m one obtains $\eta = -22$ dB and $-33.7$ dB.

In contrast, no peaks above $-50$ dB are generated by the untapered reference structure consisting of $21$ elements, each of them oscillating at $\omega_{\rm{p}}$ when alone [Fig. $2$(a)]. This dramatic difference is attributed to the strong near-field interactions between the elements of the tapered nanoantenna leading to large field enhancement in a broad spectral range [Fig. $2$(c)]. However, the strong near-field interactions between equal-size elements of the reference structure result in a redshift of its single narrowband resonance from the nominal frequency $\omega_{\rm{p}}$ \cite{bou05}.

To further quantify the rate of conversion efficiency enhancement, we also calculate $\eta$ for an idealized case of plane wave propagation in bulk ITO. We use an analytical asymptotic approximate formula that expresses $\eta$ as a function of $E_{\rm{p}}$ after the propagation distance of the plane wave $L$ \cite{agrawal,fdtd2}
\begin{eqnarray}
\eta = (3\omega_{\rm{p}}\chi^{(3)}{E_{\rm{p}}}^2L/8nc)^2
\label{eq:two},
\end{eqnarray}
where $n$ is the linear refractive index of ITO and $c$ is the speed of light. In accord with Eq. ($2$), for $E_{\rm{p}} = 4 \cdot 10^7$ V/m the conversion efficiency of $-22$ dB at $184.3$ THz can be achieved if the plane wave propagates for more than $25$ $\mu$m. By substituting the total length of the $21$-element nanoantenna ($\sim 1.7$ $\mu$m) into Eq. ($2$) we obtain $\eta_{\rm{ant}}/\eta_{\rm{bulk}} \approx 300$ [see also Fig. $3$(a)]. Note that this result does not account for linear losses in ITO.

\begin{figure}[htb]
\centerline{
\includegraphics[width=8.5cm]{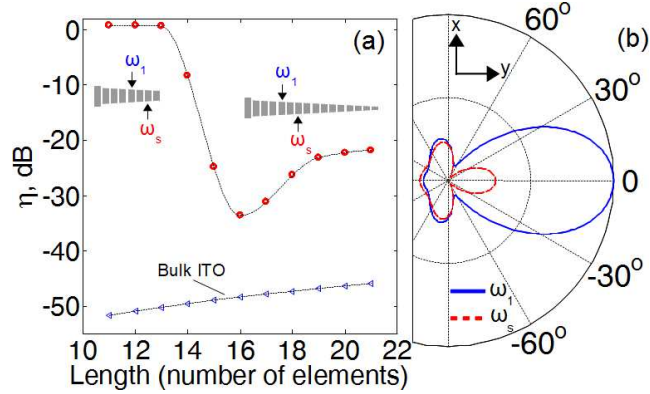}}
\caption{(a) Conversion efficiency $\eta$ as a function of the nanoantenna length (frequency $184.3$ THz, $E_{\rm{p}} = 4 \cdot 10^7$ V/m). $\eta$ for bulk ITO of the same length are given for reference. (b) Far-field power patterns of the \emph{linear} $11$-element nanoantenna at the signal frequency (red dashed line) and converted frequency $184.3$ THz (blue solid line).}
\end{figure}

Finally, we investigate the impact of the total length of the nanoantenna on $\eta$ [Fig. $3$(a)]. First, $\eta$ decreases rapidly with the total length the nanoantenna because longer architectures provide larger field enhancement. However, $\eta$ becomes anomalously large for short nanoantennas owing to uneven enhancement of the signal wave [Fig. $3$(b)] and generated wave at $184.3$ THz by the nanoantenna. It appears due to difference in of the field localization for two modes, where the element supporting the signal frequency is close to the front end of the short nanoantenna [see the insets in Fig. $3$(a)] that results in poor radiation in the forward direction. 

In conclusion, we have demonstrated a $300$-fold enhancement of the frequency conversion due to cascaded FWM in plasmonic nanoantennas. This is achieved due to judicious choice of the number and length of nanoantenna elements in accord with the frequencies involved in FWM. Apart from potential applications in nonlinear plasmonic devices, our finding can be used to enhance parametric amplification of two pre-existing waves exciting the nanoantenna at frequencies $\omega_{\rm{1}}$ and $\omega_{\rm{2}}$ .

This work was supported by the Australian Research Council. We thank D. Neshev and I. Staude for stimulating discussions.

%\pagebreak

\pagebreak

\section*{Informational Fourth Page}
%%% Complete bibliography list

\end{document}